# Efficient thermo-optic micro-ring phase shifter made of PECVD silicon-rich amorphous silicon carbide

*Li-Yang Sunny Chang* [1] *, *Steve Pappert* [2]*, and Paul K. L. Yu* [1, 2]

[1] Materials Science and Engineering Program and the Department of Mechanical and Aerospace Engineering, University of California, San Diego, 9500 Gilman Drive, La Jolla, CA 92093, USA

[2] Department of Electrical & Computer Engineering, University of California, San Diego, 9500 Gilman Drive, La Jolla, CA 92093, USA

*Corresponding author: Li-Yang Sunny Chang

 Email: lic163@eng.ucsd.edu

## Abstract

In this work, the thermo-optic coefficient (TOC) of the silicon-rich amorphous silicon carbide (a-SiC) thin film deposited by plasma-enhanced chemical vapor deposition is characterized. We found that the TOC of the film increases as its silicon content increases. A more than three-fold improvement in the TOC is measured, reaching a TOC as high as $1.88 \times 10^{-4}$ °C$^{-1}$, which is comparable to that of crystalline silicon. An efficient thermo-optic phase shifter has also been demonstrated by integrating the silicon-rich a-SiC micro-ring structure with a NiCr heater. Its tuning efficiency $P_\pi$ as low as 4.4 mW has been measured at an optical wavelength of 1550 nm. These findings make silicon-rich a-SiC a good material candidate for thermo-optic applications in photonic integrated circuits.

Silicon carbide is a material platform that attracts attention for the development of photonic integrated circuit (PIC) applications as it provides a wide range of functionalities. It possesses several properties including a wide transparency window [1], low in-waveguide propagation loss [2], as well as relatively high second-order [3] and third-order nonlinearities [4] making it an attractive material platform for the development of PICs for a wide range of applications [5]. Also, the absence of two-photon absorption at the near-infrared (NIR) wavelengths makes it advantageous over silicon devices for high-intensity nonlinear optical PICs. Several nonlinear optical devices have already been demonstrated based on silicon carbide [2, 6-8]. The special vacancy states in silicon carbide also make it a unique material for quantum photonic applications [9].

Other desirable characteristics for integrated photonic applications are the tunability and reconfigurability of the photonic circuits, which enables reconfigurable photonic devices. Many photonic applications, including optical modulators [10], and all-optical switches [11], rely on tunability to control the optical properties in PICs. A common mechanism that is used for circuit tuning is the thermo-optic effect. The efficiency of the thermal tuning is related to the materials' thermo-optic coefficient (TOC) which is a measure of the response of the material's refractive index change with the temperature. The TOC of stoichiometric silicon carbide is $2.67 \times 10^{-5}$ °C$^{-1}$ [12], which is almost one order of magnitude lower than that of silicon ($1.8 \times 10^{-4}$ °C$^{-1}$ [13]), leading to less competitive thermal tuning properties. The lower TOC and lower refractive index make stoichiometric SiC-based devices suffer from higher power consumption as well as larger device footprints.

One possible solution is to use silicon-rich a-SiC instead of stoichiometric SiC. By tuning the deposition conditions, we can change the silicon content in the material and hence design for improved optical properties. It has already been demonstrated that very high third-order nonlinearity can be obtained in silicon-rich a-SiC through the DC Kerr effect [14]. A higher refractive index facilitates a more confined mode inside the waveguide, thus allowing a smaller device footprint. In this work, we study the TOC of a-SiC thin films with different silicon contents. By measuring the resonance wavelength shift of micro-ring resonators under different temperatures, we found that, like the linear and nonlinear refractive index, the TOC can also be tuned by changing the silicon content in the a-SiC. Furthermore, we demonstrate efficient wavelength thermal tuning by using a silicon-rich a-SiC micro-ring phase shifter integrated with a NiCr heater. The results indicate that silicon-rich a-SiC provides a promising path for the development of reconfigurable linear and nonlinear PICs.

To investigate the thermo-optic properties of a-SiC with different stoichiometry, the a-SiC films are prepared by PECVD onto a thermally oxidized silicon substrate. By varying the flow ratio of the precursors, silane (SiH4) and methane (CH4), a-SiC films with different stoichiometry can be realized. The refractive index of the films ranges from 2.48 to 2.8 at 1550 nm as measured by the ellipsometer. Table 1 summarizes the materials index and the flow rate of precursor gases for the three films SRC1, SRC2, and SRC3 investigated in this work. Energy-Dispersive X-ray spectroscopy (EDX) measurement in our previous work indicates that the silicon content increases as the SiH4/CH4 gas flow ratio increases [14]. The EDX estimated silicon content of these samples is shown in Table 1.

To determine the materials' TOC, we fabricate three different sets of micro-ring resonators using the three different films (SRC1, SRC2, and SRC3) we prepared earlier. The a-SiC thin film is patterned by electron beam lithography (EBL) and the patterns are transferred by reactive ion etching (RIE) with CHF3 and O2 etching chemistry. The samples are then encapsulated by a silicon dioxide cladding layer using PECVD. Finally, the input and the output of the waveguides are polished by focused ion beam (FIB) milling to ensure better edge coupling efficiency. Figure 1(a) illustrates the fabrication process for the waveguides. Figure 1(b) shows the Scanning Electron Microscopic (SEM) cross-section image of the waveguides. The waveguide width and height are 875 nm and 320 nm respectively, which are designed to ensure single-mode propagation in the waveguide as simulated by a Finite Difference Eigenmode (FDE) solver built in Lumerical MODE. Figure 1(c) shows the FDE simulated mode profile of the waveguide made on SRC3.

The transmission spectra of the ring resonators are obtained using a fiber-in-free-space-out setup shown in Figure 2. The output from a wavelength-tunable CW laser ranging from 1465 to 1575 nm is applied through a single-mode fiber, an Erbium-Doped Fiber Amplifier (EDFA), and a polarization controller to make sure that only the TE-polarized mode is excited in the waveguide. A lensed fiber is used to edge couple the laser light to the device under test (DUT). The output spectrum is filtered by a TE-mode crystal polarizer and then collected using a large-area InGaAs photodetector. The DUT is mounted onto a stage that is temperature controlled by a thermo-electric cooler (TEC) module with an accuracy of ± 0.1 °C.

We measure the transmission resonance wavelength shift by changing the TEC temperature. Figure 3(a) shows the transmission spectra near the 1550 nm wavelength measured from the SRC3 ring resonators at temperatures ranging from 20 °C to 35 °C. The red shift of the resonance wavelength with temperature indicates a positive TOC in a-SiC. We apply the Lorentzian fit function to the resonance response and then calculate the resonance wavelength. The resonance wavelength shifts versus temperature are plotted in Figure 3(b). We can see that SRC3 has the largest resonance shift in response to the temperature change. Hence, we demonstrated that larger silicon content leads to a larger temperature response.

The TOC of the materials can be extracted from the resonance shift. Temperature dependence of the resonance wavelength can be related to the effective index $n_{eff}$ as [15]:

$$\frac{d\lambda}{dT} = \left(n_{eff} \cdot \alpha_{sub} + \frac{dn_{eff}}{dT}\right)\frac{\lambda}{n_g} \qquad (1)$$

where $\alpha_{sub}$ is the thermal expansion coefficient of the silicon dioxide, $\lambda$ is the resonance wavelength, and $n_g$ is the group index of the waveguide. The change of the effective index is due to the interaction of the mode with the waveguide core and cladding and can be related to the change of the materials' index as [16]:

$$\frac{dn_{eff}}{dT} = \Gamma_{SiO2}\left(\frac{dn}{dT}\right)_{SiO2} + \Gamma_{a-SiC}\left(\frac{dn}{dT}\right)_{a-SiC} \qquad (2)$$

where $\Gamma$ is the overlap integral of the SiC waveguide core and the silicon dioxide cladding and $dn/dT$ is the TOC of the materials. Figure 3(c) shows the extracted TOC of the three films investigated. A clear trend of increasing the materials' refractive index leading to an increase in the TOC is observed. The result shows a more than three-fold improvement in the TOC from SRC1 ($5.78 \times 10^{-5}\ °C^{-1}$) to SRC3 ($1.88 \times 10^{-4}\ °C^{-1}$). The highest TOC achieved in SRC3 is comparable to that of crystalline Si.

The SRC3 micro-ring is further investigated by integrating it with a NiCr heater to form a thermo-optic phase shifter. The NiCr heater film is positioned on top of the a-SiC waveguide. By applying current through the heater, it can effectively heat the waveguide underneath. The simulation of temperature distribution in the devices is done by the Lumerical Heat solver. Figure 4(a) shows the simulated temperature distribution with 20 mW of applied power. The NiCr heater is designed to be 2 μm wide, 200 nm thick, and has a 2 μm separation from the silicon-rich a-SiC waveguide. From the simulation, the temperature in the waveguide is increased by 1.319 °C per 1 mW of applied power, which is used in Figure 5 (b) in simulating the wavelength shift with respect to applied power. The NiCr heater is formed using the lift-off process after the fabrication steps shown in Figure 1(a). It is patterned using NR9–3000 negative resist and Heidelberg MLA150 maskless aligner. A 200 nm layer of NiCr is then deposited by the Denton 18 sputter system. Finally, a 300 nm gold contact pad is formed by the same lift-off process. The resistance of the NiCr heater is measured to be 3600 Ω using the B1500 Agilent Semiconductor Analyzer. Figure 4(b) shows the optical microscope (OM) image of the top view of the fabricated micro-ring heater and the gold contact pad which illustrate the heating area.

The thermo-optic phase shifter is then measured by applying different powers and collecting the transmission spectra using the same setup depicted in Figure 2. Figure 5(a) shows the measured resonance wavelength around 1552 nm with different applied powers. We extract the resonance wavelength shift and plot it versus applied power in Figure 5(b). The simulated results are also shown for comparison. The simulation data is collected by simulating the device temperature under different applied powers and calculating the effective index through the FDE model. The transmission spectra of the phase shifter are calculated using MATLAB and the resonance shift can be extracted from the transmission spectra under different power levels. From Figure 5(b), one can see a good agreement between the experimental results and the calculated results. A linear trend is observed, and a 0.11 nm/mW of thermal tunability is estimated from the slope of the linear fitting of the data.

The power requirement for π phase shift Pπ for the resonator-based thermo-optic switches can be defined as the power consumption to tune the resonator by a phase shift of the product of π and the full width at half maximum (FWHM) of the resonance peak and can be calculated as [17]:

$$P_\pi = \frac{\text{FWHM} \times \pi}{\eta} \qquad (3)$$

where η is the thermal tunability. The FWHM for our device is 0.155 nm and the $P_\pi$ is calculated to be 4.4 mW. The small $P_\pi$ indicates the potential for the realization of many efficient thermo-optic integrated circuit applications.

In summary, we have measured the TOC in a-SiC thin films as the silicon content increases. We find that by increasing the silicon content, we can effectively increase the TOC of the material. More than three-fold improvement is achieved in our silicon-rich a-SiC thin films compared with the stoichiometric a-SiC thin film. The highest TOC we get is $1.88 \times 10^{-4}\ °C^{-1}$, which is comparable to that of crystalline silicon. A highly efficient thermo-optic phase shifter is demonstrated by integrating a NiCr heater with a silicon-rich a-SiC micro-ring resonator. A thermal tunability of 0.11 pm/mW is achieved in the micro-ring resonator which translates to a $P_\pi$ of 4.4 mW. With the high TOC and easy tuning of the material properties, we believe that silicon-rich a-SiC provides a promising materials platform for reconfigurable PIC applications.

## Acknowledgments

The authors thank members of UCSD Nano3 cleanroom especially Dr. Maribel Montero, for their expertise in sample preparation and fabrication. This work was supported by the Office of Naval Research (N00014-18-1-2027).

## Disclosures

The authors declare no conflicts of interest.

## Data Availability

The data that support the findings of this study are available from the corresponding author upon reasonable request.

**Table 1. Refractive index n at λ=1550 nm and the precursor gasses flow rate of the three a-SiC films SRC1, SRC2, and SRC3 in this study.**

| Materials | n | SiH4 flow rate(sccm) | CH4 flow rate (sccm) | Si content (%) |
|---|---|---|---|---|
| SRC1 | 2.48 | 200 | 100 | 61 |
| SRC2 | 2.61 | 200 | 50 | 70 |
| SRC3 | 2.80 | 300 | 50 | 80 |

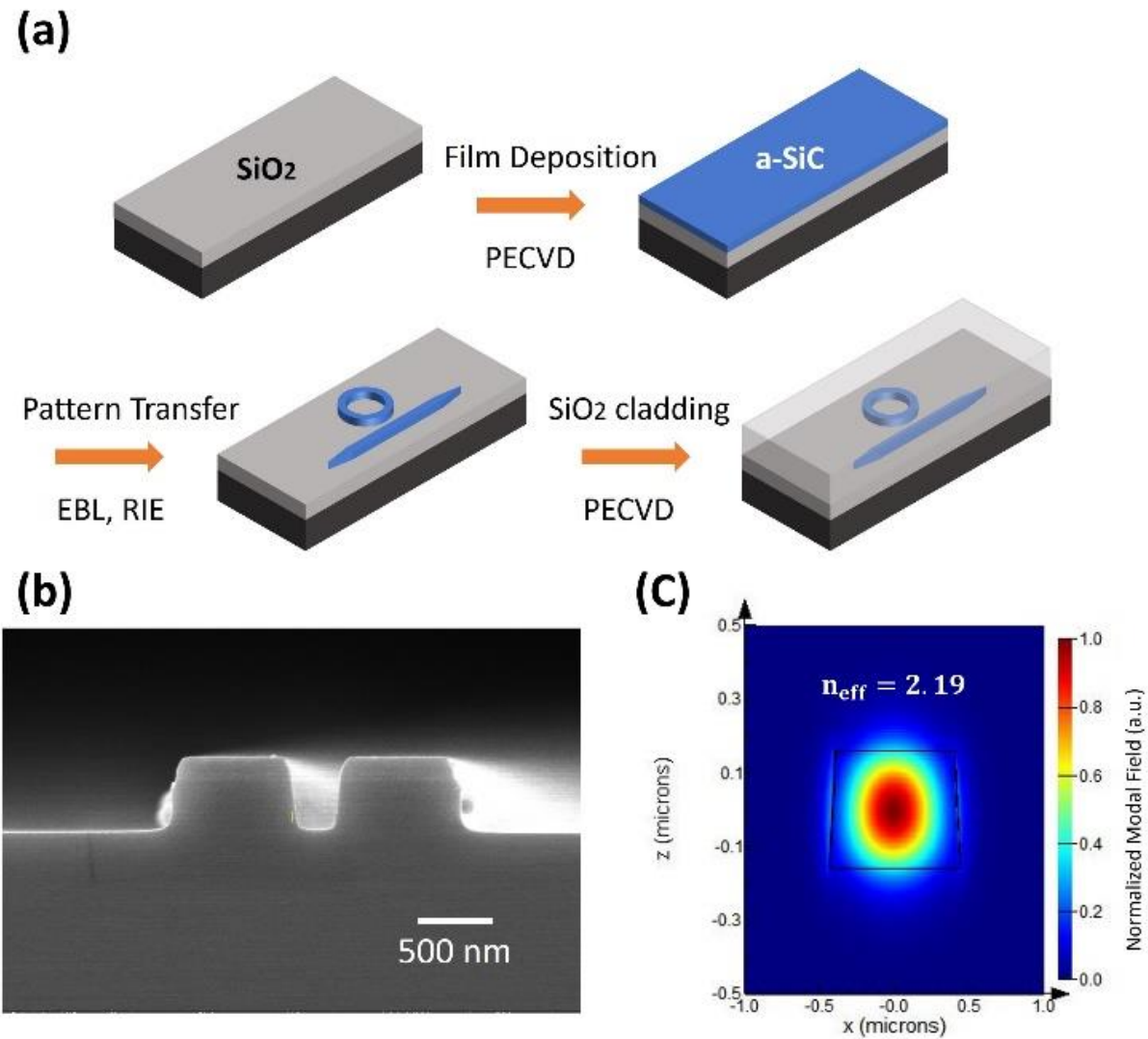

**Figure 1** (a) Schematic of the fabrication process of the silicon-rich a-SiC micro-ring resonators. (b) The SEM cross-section image of the waveguide. (c) The simulated mode profile for the waveguide made on SRC3. The simulated effective index is 2.19.

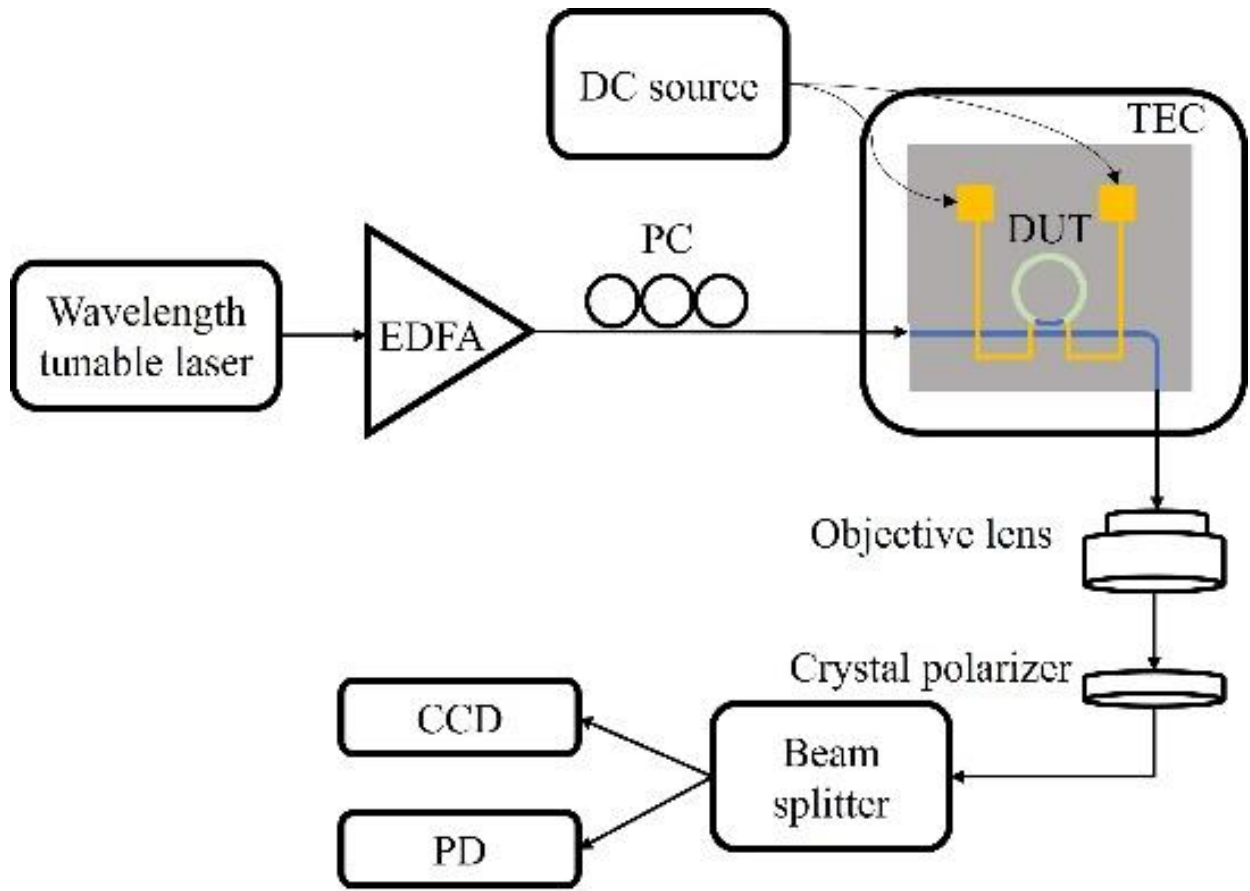

**Figure 2** Schematic of measurement setup for thermal-optic characterization.

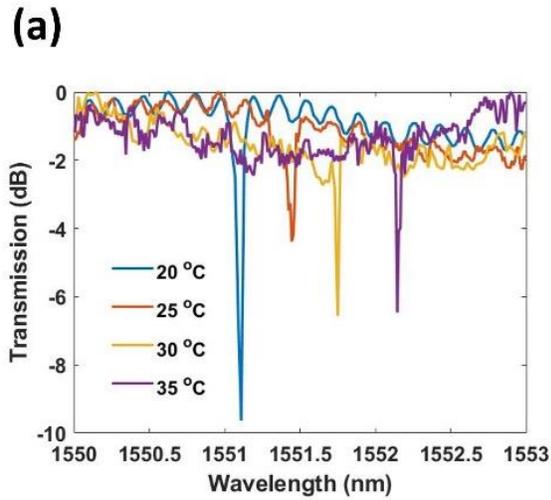

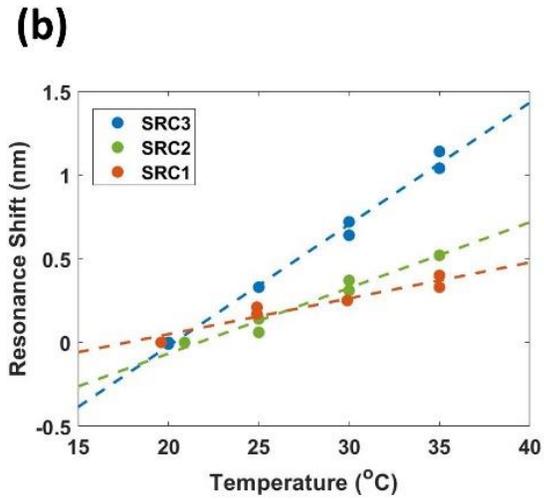

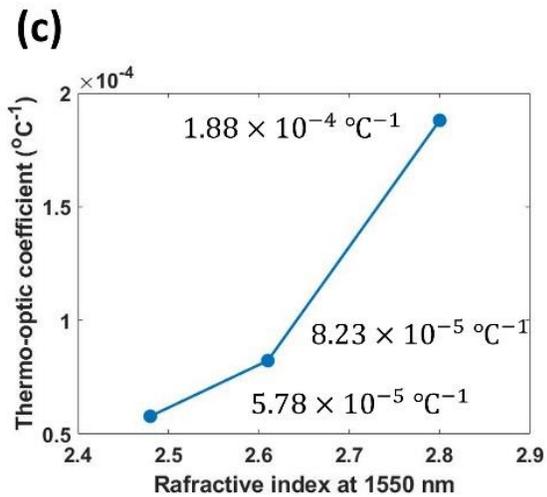

**Figure 3** (a) Transmission spectra of the micro-ring resonator made by SRC3 under 20 °C, 25 °C, 30°C, and 35 °C at a wavelength around 1550 nm. (b) Extracted resonance wavelength shift for the micro-ring resonators with three different stoichiometry SRC1, SRC2, and SRC3. (c) Extracted thermo-optic coefficient for the three films.

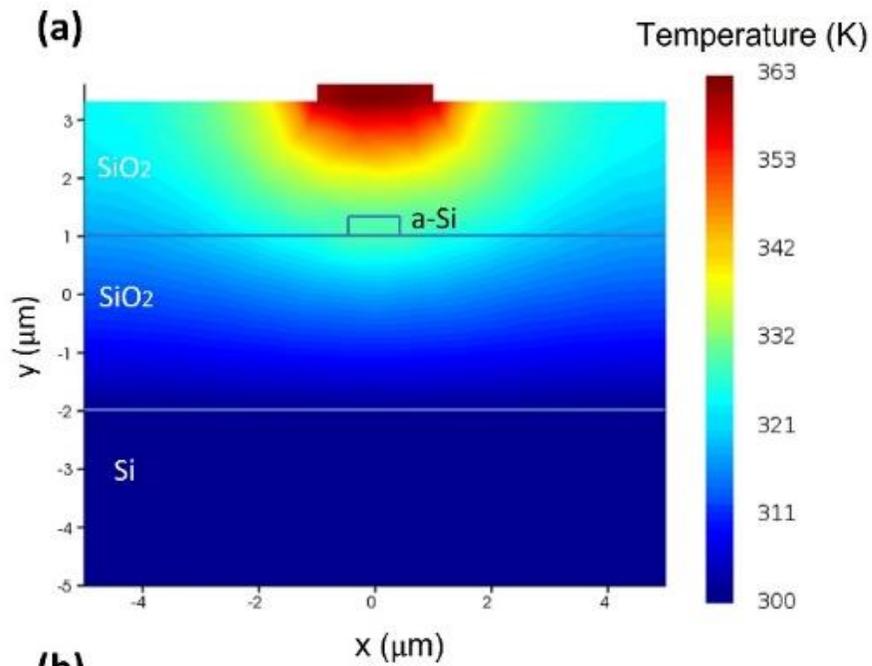
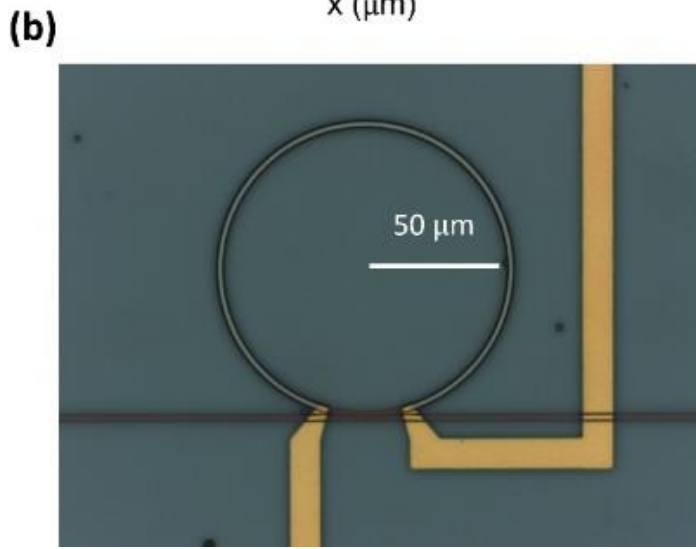

**Figure 4** (a) Temperature distribution simulation for the a-SiC micro-ring resonator heated by NiCr heater under 20 mW of applied power. (b) The OM top view of the fabricated micro-ring phase shifters.

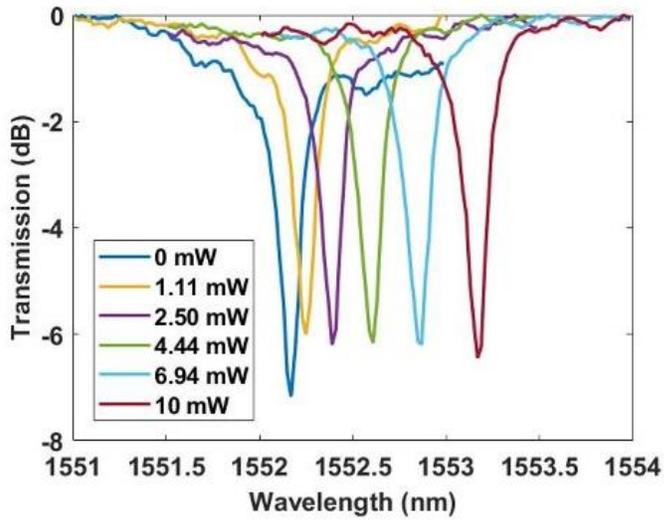

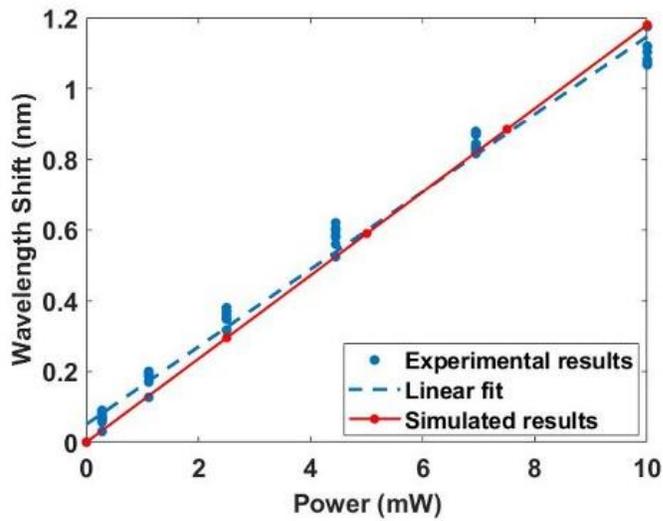

**Figure 5** (a) The transmission spectra of the micro-ring phase shifter made from SRC3 under different applied voltages at a wavelength around 1552 nm. (b) The simulated and the experimental result for the thermal tunability of the phase shifter. The extracted thermal tunability is 0.11 nm/mW.